\documentstyle[11pt,paspconf,psfig,epsf]{article}

\begin{document}

\title{Selection Effects at 21cm}
\author{F.H. Briggs}
\affil{Kapteyn Astronomical Institute, P.O. Box 800, 9700 AV
  Groningen, The Netherlands }

\begin{abstract}
Surveys in the 21cm line of neutral hydrogen are testing the
completeness of the catalogs of nearby galaxies.  The remarkable
observational fact is that the potential wells that confine gas to sufficient
density that it can remain neutral in the face of ionizing radiation
also provide sites for star formation, so that there are no known cases
of neutral intergalactic clouds without associated star light. 
\end{abstract}


\section{Introduction}

Invisible hydrogen clouds are common place in the vicinity of
galaxies.  This ``low surface brightness'' material lies
in tidal debris from galaxy interactions (cf. Haynes et al 1979,
van der Hulst 1979, Yun et al 1994,
 Putnam et al 1998), in the extended HI disks
that facilitate kinematical studies (cf Broeils \& van Woerden 1994), 
in outlying gas rings around
galaxies and pairs of galaxies
 (Schneider 1989, van Driel et al 1988), in the
high velocity clouds around the Milky
Way (Wakker and van Woerden 1998), and in a number of other structures
for which the nature of the extended LSB gas is less clear
(cf. Fisher \& Tully 1976, Simkin et al 1987, Giovanelli \& Haynes 1989).

Despite vigorous searching, no examples of isolated neutral hydrogen
clouds of a truly intergalactic nature have been discovered, and, in fact,
for some years, the upper limit to the baryon content of intergalactic
clouds with HI masses in the range comparable to normal galaxies has
been known to be cosmologically insignificant (Fisher \& Tully 1981).
This finding is in sharp contrast to the recognition that the {\it ionized}
intergalactic medium may form an important, perhaps dominant, 
reservoir of the Universe's
baryons (Rauch et al 1997, Shull et al 1996).

What selection effects influence and limit the neutral hydrogen picture of the
nearby universe?  What properties could neutral gaseous systems have
that would allow them to escape detection in the surveys that have been made 
so far?  

\section{Selection Effect `0': the need for neutrals}

Clearly, a easy way to hide hydrogen from observers at 21cm 
wavelength is to ionize it.  However, 
even ionized clouds will have some
neutral fraction, and in principle, long integrations with
sensitive receivers could detect the 21cm emission from 
predominantly ionized clouds. The optical depth in the 21cm
line of a cloud with neutral $H^o$ column density $N_{HI}$,
velocity width $\Delta V$ (FWMH) and excitation temperature
$T_{spin}$ is
\begin{equation}
\tau_{21} \approx \left(\frac{N_{HI}}{2{\times}10^{21}\;{\rm cm}^{-2}}\right)
            \left(\frac{10\;{\rm km\;s}^{-1}}{\Delta V}\right)
            \left(\frac{100\;{\rm K}}{T_{spin}}\right)
\label{tau.eqn}
\end{equation}
Under the conditions for which emission is normally observed from HI
clouds (i.e., the excitation temperature is significantly above the
temperature provided by the background radiation flux 
($T_{spin} \gg 2.7\;{\rm K}$) and $\tau_{21} \ll 1$), the brightness
temperature of the emergent radiation from a cloud of 
neutral $H^o$ column density $N_{HI}$ is
\begin{equation}
T_B \approx \tau_{21}T_{spin} 
    \approx \left(\frac{N_{HI}}{2{\times}10^{19}\;{\rm cm}^{-2}}\right)
            \left(\frac{10\;{\rm km\;s}^{-1}}{\Delta V}\right) \;{\rm K}
\label{tb.eqn}
 \end{equation}
Under these conditions, the integration time required for a 
radio telescope to detect a cloud 
(with $\Delta V\approx 20\;{\rm km\;s}^{-1}$)
that fills the telescope beam is approximately 
one second for $N_{HI}=2{\times}10^{19}$, 
two minutes for $N_{HI}=2{\times}10^{18}$, and
three hours for $N_{HI}=2{\times}10^{17}\;{\rm cm}^{-2}$.
The latter column density is of interest because clouds with
$N_{HI}\leq 3{\times}10^{17}\;{\rm cm}^{-2}$ are optically thin
for wavelengths shortward of the Lyman limit ($\tau_{LL}\leq 1$)
and are therefore vulnerable to ionizing radiation with the expectation
that the hydrogen within them will be predominantly ionized. For example,
it has long been expected that photoionization will cause
the outer edges of gaseous galaxy
disks to show abrupt declines in the column density of neutral
atoms (Sunyaev 1969, Maloney 1993, Corbelli \& Salpeter 1993);
deep integrations in the outskirts of galaxies and in the
intergalactic medium at large, may detect local analogs
of the highest column density
clouds in the Lyman-$\alpha$ forest and Lyman-limit population
identified through quasar absorption-line studies (cf. Hoffman et al 1993,
Charlton et al 1994).

Even neutral gas can be hidden from 21 cm emission-line surveys if the 
excitation temperature $T_{spin}$
is low, as might occur if the gas density is so low that $T_{spin}$
is not coupled to the gas kinetic temperature $T_k$ or if $T_k$ is
intrinsically low.  As indicated by Eq.~\ref{tau.eqn},
such low $T_{spin}$ clouds would be
good absorbers, which should appear in absorption against
background continuum sources (cf. Corbelli \& Schneider 1990).
Of course, in instances when the 21 cm line is optically thick, 
application of Eq.~\ref{tb.eqn} will underestimate the column 
density and the HI cloud masses
(cf. Dickey et al 1994, Braun 1995, Braun 1997).

\section{Selection effect 1: `optical selection' of targets}

An historically important selection effect is that the vast
majority of extragalactic 21cm line observations have been made
with telescopes pointed directly at optically selected objects.
Only within the past decade has it become technically feasible
to make ``optically blind'' surveys of sufficiently large volumes
of space with adequate sensitivity to recover the known galaxy
population, let alone identify new populations of intergalactic
clouds or gas-rich LSB galaxies.  However, if there were such populations
containing HI masses comparable to those in ordinary galaxies and 
comparable in spatial number density,
then the historically important attempts (Shostok 1977,
Materne et al 1979, Lo \& Sargent 1979, Haynes \& Roberts)
would have discovered them. The lack of detections in these surveys,
combined with the absence of serendipitous detections in the studies
targeted on optically selected galaxies, led Fisher and Tully (1981)
to conclude that the mass in intergalactic HI clouds in 
the range $10^7$ to $10^{10}M_{\odot}$ could only be a small fraction of the 
total mass in galaxies, and thus can only contribute a tiny
fraction of the total cosmological mass density. 
Briggs (1990) performed an update of this
sort of analysis to conclude that the amount of the HI mass contained in
un-identified new populations is only a few percent of HI mass contained
in catalogued galaxies over this same mass range.

There have been a few surprises
during this period of extensive observation of optically
selected targets,  including discoveries of uncatalogued
HI-rich clouds of undetectably low optical surface brightness 
(cf. Fisher \& Tully 1976, Schneider 1989, Giovanelli \&
Haynes 1989). However, in every case, the neutral gas is clearly associated
in position and redshift velocity with optically visible galaxies,
implying that the gravitational potential belonging to the optical
galaxy is responsible for the confinement of the hydrogen to sufficient
density that it remains neutral.

Similar conclusions result from the recent blind surveys that have
been successful at compiling samples of HI-selected galaxies and recovering
the known late-type galaxy population (cf. Szomoru et al 1994,
Henning 1995, Zwaan et al 1997, Spitzak \& Schneider 1998, Kraan-Korteweg
et al 1998).  All the confirmed HI detections from these surveys
are found to be associated with optical emission from star light,
provided that the objects are located at sufficiently high galactic
latitude that they are not heavily obscured and are not located
close to a bright foreground star.  This finding lends credence to
the idea that a reasonably complete picture of the neutral gas content of
the nearby universe can be obtained from a study of the HI gas in
the large optically galaxy samples (cf. Rao \& Briggs 1993, Briggs 1997).

\section{Selection effects for spectroscopic features}

Surveys that require the detection of spectroscopic emission features
differ in several respects from flux- or magnitude-limited surveys.
These differences are illustrated in Fig.~1 and discussed in the
following paragraphs. 

Until recently, the limitations to
radio spectrometer bandwidth and the number of available spectral
channels within the spectrometer have often caused surveys to be limited in
redshift depth, 
since they were unable to cover large redshift ranges efficiently.
Most of these surveys have sufficient sensitivity to detect
large HI masses of $M_{HI}> 10^{10}$M$_{\odot}$  beyond this
restricted survey depth.  At the low mass extreme $M_{HI}<10^7$M$_{\odot}$
nearly all surveys are limited in depth by sensitivity, since integration
times become very long to detect these tiny masses at distances of order
10~Mpc or more.

\begin{figure}
\psfig{figure=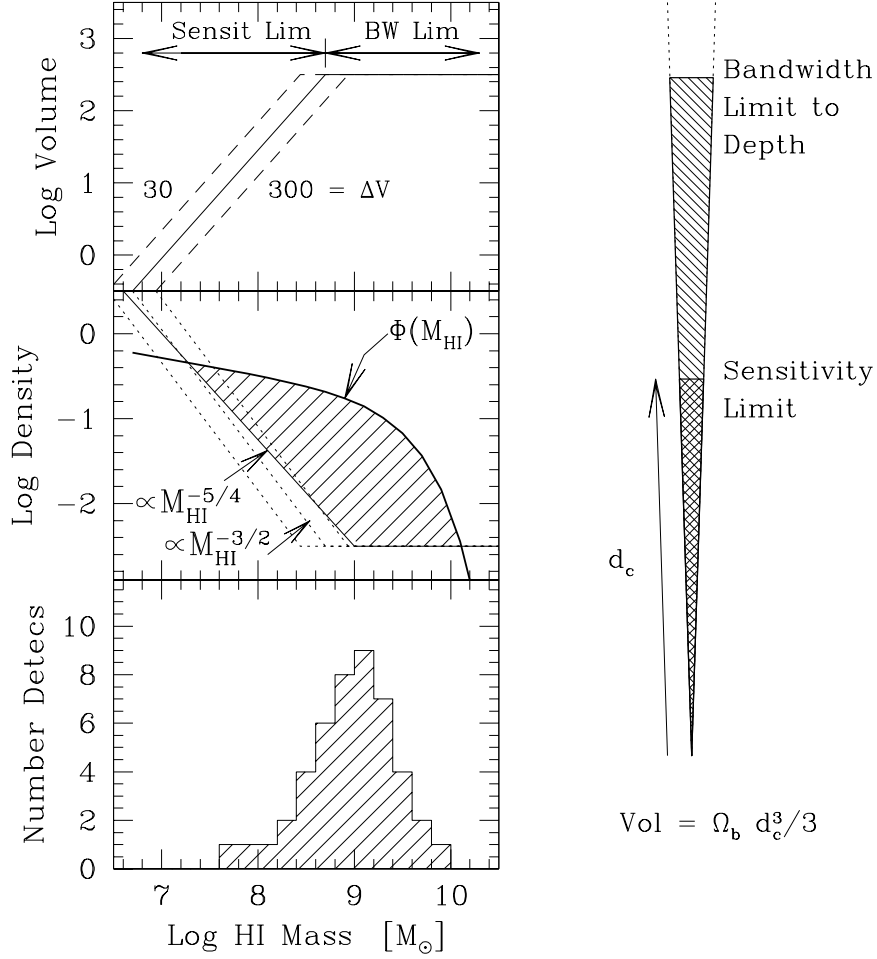,width=14cm,angle=0}
\caption{21cm Line Detection Sensitivity. {\it Right:} The volume probed
by a radio telescope beam depends on the beam solid angle $\Omega_b$
and the depth $d_c$ to which a given HI mass can be detected.  For
large masses, the detection volume is often limited by the spectrometer
bandwidth. {\it Left top:} In the ``sensitivity limited regime'' applicable
for small
HI masses, the survey volume also depends on velocity width $\Delta V$
shown here for $\Delta V =$ 30, 100 and 300 km~s$^{-1}$.
{\it Left middle:} The quantity $({\rm survey\; volume})^{-1}$ specifies a
sensitivity function to objects whose space density is $\Phi(M_{HI})$,
here given as number per volume per decade of mass. In the low-mass
(sensitivity-limited) regime, including the
trend of increasing rotation speed with mass implies a sensitivity
function $\propto M_{HI}^{-5/4}$ rather than the
$M_{HI}^{-3/2}$ dependence for a purely ``flux 
limited'' survey. {\it Left bottom:}
Number of detections expected in bins of 0.2 Dex.
}
\label{sens.fig}
\end{figure}

The velocity width of spectral features also  influences the detection
efficiencies. If two galaxies with the same HI mass $M_{HI}$ 
are compared, the one with the narrower profile will be easier
to detect; the narrower profile could be detected to greater
distance. 
If all galaxies had the same velocity spread, then the limiting distance
to which each galaxy could be detected $d_c$ would be
$\propto M_{HI}^{1/2}$, according to the inverse square law, and
the volume within which each could be detected would be
proportional to $M_{HI}^{3/2}$. The noise level in flux density, $\sigma$,
in a spectrum that has been optimally
smoothed to match the profile is 
$\sigma = \sigma_o\sqrt{\Delta v_o/\Delta V}$ for a spectrum originally
recorded with channel spacing (resolution) of $\Delta v_o$ for which
the noise level is $\sigma_o$.
The HI mass for a spectral
feature of strength $S_{Jy}$ is computed from the integral
over the profile,
$M_{HI}= 2\times 10^5 d_{Mpc}^2 \int S_{Jy} dV_{km/s}M_{\odot}$. Thus,
the minimum detectable HI mass is 
$M_{HI} \propto 5\sigma \Delta V d^2 \propto d^2\sqrt{\Delta V}$,
if the minimum detectable profile is modeled as a rectangle of height
in flux density $\Delta S = 5\sigma$ and width $\Delta V$.
A sort of HI Tully-Fisher relation (cf. Briggs \& Rao 1993) has 
$\Delta V \propto M_{HI}^{1/3}\sin i$ for galaxies with inclination $i$
relative to the plane of the sky,
leading to the result that $d_c \propto M_{HI}^{5/12}\sin^{-1/4}i$.
Note that the $\sin^{-1/4}i$ factor is substantially
different from unity for a only small fraction of a randomly oriented sample.
The net result is that the
survey volume in which a HI mass would be detected is more closely
$volume\propto M_{HI}^{5/4}$ in the sensitivity-limited regime.  Fig.~1
illustrates the difference in survey volumes for profiles of width
30, 100 and 300 km~s$^{-1}$. These volumes are translated into 
survey ``sensitivity functions'' ($\propto 1/volume$) in units of
Mpc$^{-3}$ for comparison with HI mass functions; if a survey finds no
masses of $10^7M_{\odot}$ in a volume of 1~Mpc$^3$, then an estimate
of the upper limit to the density of $10^7M_{\odot}$ objects 
is $\sim$1~Mpc$^{-3}$.

Figure 1 (left middle panel) illustrates a typical survey mass sensitivity
in comparison to an HI mass function $\Phi(M_{HI})$ plotted as number of
objects per
Mpc$^{-3}$ per decade of mass. For mass ranges where
the survey sensitivity function lies below the mass function, 
it become probable that the survey will detect galaxies of that mass.
The shaded area represents the mass range where detections should occur
in the number as indicated in the lower plot, where detections are binned
in 0.2 decade bins. In the regions below ${\sim}10^{7.3}$ and 
above  ${\sim}10^{10.1}$, this survey would probably obtain no detections
but could place upper limits on the space density of galaxies in these mass
ranges.

The knowledge of the space density of tiny HI-rich extragalactic objects
has remained highly uncertain because of the difficulty in surveying
a sufficiently large volume with adequate sensitivity to detect them.
A further complication is that the small masses have narrow velocity 
widths without the distinctive double horned profiles that characterize
large spiral galaxies with flat rotation curves; this means that
low mass systems can be more easily confused with radio interference.

The poor diffraction-limited angular resolution provided by most
single-dish radio telescopes gives rise to confusion problems,
and many detections in blind surveys may be multiple galaxy systems.
Small galaxies in close proximity to large ones can easily be missed,
if their redshift velocities fall within the range spanned by the
emission profiles of bright, dominant galaxies.
  
\subsection{Computation of the HI-mass function}
 
\begin{figure}
\hglue 2.0cm\psfig{figure=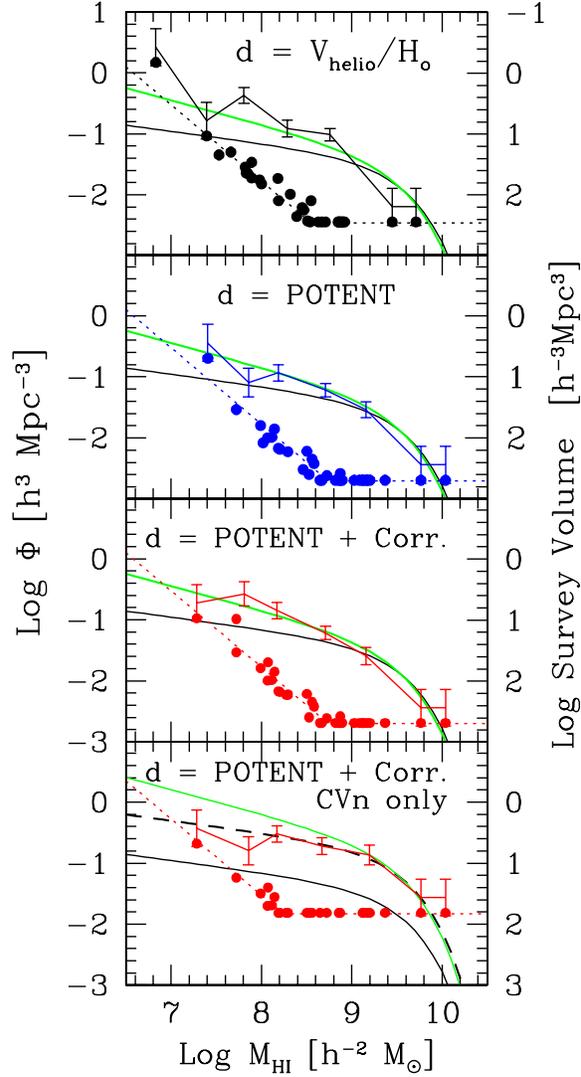,width=8.5cm,angle=0}
  \caption[]{HI-mass function for the Canis Venatici survey
volume (Kraan-Korteweg et al 1998), 
normalized to number of objects per decade of mass.
The smooth solid curve is
the analytic form derived by Zwaan et al (1997) with a slope
of $\alpha =-1.2$, the grey line has a slope of $\alpha=-1.4$
(Banks et al 1998). The bottom panel
shows the result restricted to the CVn-group regions ($<$1200 km~s$^{-1}$),
and the dashed curve represents the Zwaan et al HI-mass 
function multiplied by a factor of 4.5. The dotted line in each panel gives 
an indication of the volume probed as a function of mass 
(see right vertical axis); the points give 1/V$_{max}$ 
for each of the galaxies in the sample, taking into account 
the different velocity widths.}
\label{mfnct_quad.fig}
\end{figure}
 
A number of subtle effects enter when the space density of
gas-rich galaxies, the HI-mass function, is constructed.
These problems are especially acute for the small HI masses, which
can only be detected in blind surveys if the objects are very close
by. Some of these problems, including the effects of peculiar velocities,
deviations from pure Hubble flow, and large-scale fluctuations in
density, are illustrated in  
Fig.~\ref{mfnct_quad.fig}, which presents the recent Nan\c{c}ay
optically-blind, 21cm line survey of the Canis Venatici region.
First, the top panel shows the number density of galaxies as computed using
distances, HI masses, and sensitivity volumes based on conversion
of heliocentric velocities to distance $V_{hel}/H_o$. The mass functions are
binned into half-decade bins, but are scaled to give number of
objects per decade. The value for each decade is computed from 
the sum $\Sigma\; 1/V_{max}$, where V$_{max}$ is the volume of the
survey in which a galaxy with the properties $M_{HI}$ and $\Delta V$
could have been detected. A steeply rising, low mass tail results from this
calculation due to one galaxy, UGC~7131,  which  
is is treated in this naive calculation as a very nearby, but low mass
object.  Placed at the greater distance implied by independent
distance measurement (Makarova et al 1998), it becomes
more massive, and it finds its place among  
other galaxies of greater velocity
width and higher HI mass in the higher mass bins, as shown in the
lower panels.

An improved calculation based on POTENT distances,
which include local deviations from uniform Hubble expansion 
(Bertschinger et al 1990),  is displayed
in the second panel. In the third panel, four galaxies from
the  Nan\c{c}ay sample
with independent distance measurements have been plotted 
according to their revised distances.
In the 4th panel we have restricted our sample to include only
the overdense foreground region that contains the CVn and Coma
groups, i.e. the volume within $V_{hel} < $1200  km~s$^{-1}$ and about 
1/2 the RA coverage (about half the solid angle) of the full survey.
Large HI-mass galaxies can be detected throughout the volume we surveyed, but 
small galaxies can be detected only in the front part of our volume.
The volume normalization factors, which are used to compute the mass
function, are sensitivity limited for the small masses to only the
front part of our survey volume.  For the large masses, the V$_{max}$'s 
include the whole volume, including the volume where the numbers of
galaxies are much less. Hence, when restricting the ``survey volume'' 
we get a fairer comparison of the  number of little galaxies to the 
number of big ones.  

In all four panels the solid line represents the HI-mass function
with a slope of $\alpha=-1.2$ as derived by Zwaan et al (1997), 
whereas the grey line represents
an HI-mass function with a slope of $\alpha=-1.4$ as deduced by
Banks et al (1998) for a survey in the
CenA-group region. 
Restricting our volume to the dense foreground region
including ``only'' the CVn and Coma groups, we find that 
the Zwaan et al HI-mass function with a scaling factor of 4.5
to account for the local overdensity (dashed line in the bottom panel)
gives an excellent fit to the data.

Clearly the computation of mass functions for tiny dwarfs 
will remain vulnerable to
subtleties, due to the small volumes in which they can be detected
and the accompanying uncertainties in distance.

\section{Conclusion: physical selection}

The largest reservoirs of HI gas in the nearby universe are found in
large galaxies. The bulk of the HI associated with low optical surface
brightness falls in the outskirts of large galaxies. Apparently an
additional consequence of the gravitational confinement that 
preserves gas neutrality is the production of
stars, since no cases of free-floating HI clouds away from galaxies
have been found.  Deep integrations in the 21cm line 
over long sightlines should
eventually detect the highest column density clouds of the
ionized Lyman-$\alpha$ forest.



\begin{references}
\reference Banks, G.D., Disney, M.J., Knezek, P., et al 1998, in prep
\reference Bertschinger, E. et al 1990, ApJ, 364, 370
\reference Braun, R. 1997, ApJ, 484, 637 
\reference Braun, R. 1995, BAAS, 187, 6501 
\reference Briggs, F.H., \& Rao, S. 1993, ApJ, 417, 494
\reference Briggs, F.H. 1997, ApJ, 484, 618
\reference Broeils, A.H., van Woerden, H. 1994, A\&AS, 107, 129
\reference Charlton, J.C., Salpeter, E.E., \& Linder, S.M. 1994, ApJ, 430, L29
\reference Corbelli, E., \& Salpeter, E.E. 1993, ApJ, 419, 104
\reference Corbelli, E., \& Schneider, S.S. 1990, ApJ, 356, 14
\reference Dickey, J.M., Mebold, U., Marx, M., et al 1994, A\&A, 289, 357
\reference Fisher, J.R., \& Tully, R.B. 1976, A\&A, 53, 397
\reference Fisher, J.R., \& Tully, R.B. 1981, ApJ, 243, L23
\reference Giovanelli, R. \& Haynes, M.P.  1989, ApJ, 346, L5
\reference Haynes, M.P., \& Roberts, M.S. 1979, ApJ, 227, 767
\reference Haynes, M.P., Giovanelli, R., \& Roberts, M.S. 1979, ApJ, 229, 83
\reference Henning, P.A. 1995, ApJ, 450, 578
\reference Hoffman, G.L., Lu, N.Y., Salpeter, E.E., et al 1993, AJ, 106,39
\reference Kraan-Korteweg, R.C., van Driel, W., Briggs, F.H.,
Binggeli, B., \& Mostefaoui, T.I. 1998, A\&A, in press
\reference Lo, K.Y., \& Sargent, W.L.W. 1979, ApJ, 227, 756
\reference Maloney, P. 1993, ApJ, 414, 41
\reference Materne, J., Huchtmeier, W.K., \& Hulsbosch, A.N.M. 1979, MNRAS, 18, 563
\reference Putnam, M.E. Gibson, B.K., Staveley-Smith, L., et al 1998,
Nature, 394, 742
\reference Rao, S.M., \& Briggs, F.H. 1993, ApJ, 419, 515
\reference Rauch, M., et al 1997, ApJ, 489, 7
\reference Schneider, S.E. 1989 , ApJ, 343, 94
\reference Shostok, G.S. 1977, A\&A, 54, 919
\reference Shull, J.M., Stocke, J.T., \& Penton, S. 1996, AJ, 111, 72
\reference Simkin, S.M., Su, H-J., van Gorkom, J., Hibbard, J. 1987,
  Sci, 235, 1367
\reference Spitzak, J., \& Schneider, S.E. 1998, ApJS, in press
\reference Sunyaev, R.A. 1969, Astrophys.Lett., 3, 33
\reference Szomoru, A., Guhathakurta, P., van Gorkom, J.H., Knapen, J.H.,
Weinberg, D.H., \& Fruchter, A.S. 1994, AJ, 108, 491
\reference van der Hulst, J.M. 1979, A\&A, 75, 97
\reference van Driel, W., van Woerden, H., Schwarz, U.J., \&
 Gallagher, J.S. 1988, A\&A, 191, 201
\reference Wakker, B.P., \& van Woerden, H., 1997, ARA\&A, 35, 217
\reference Yun, M.S., Ho, P.T.P., \& Lo, K.Y. 1994, Nature, 372, 530
\reference Zwaan, M.A., Briggs, F.H., Sprayberry, D., \& Sorar, E.
1997, ApJ, 490, 173
\end{references}
\end{document}